\documentclass[aps,pra,twocolumn,groupedaddress,showpacs]{revtex4}
\usepackage{latexsym}
\usepackage{amsmath}
\usepackage{graphics}
\begin{document}
\title{Low-energy modes of spin-imbalanced Fermi gases in BCS phase}
\author{F. Matera$^{1}$}
\email{matera@fi.infn.it}
\author{M. F. Wagner$^{2}$}
\email{mfwagner@mailaps.org}
\affiliation{
$^1${\small\it Dipartimento di Fisica e Astronomia, Universit\`a degli Studi
di Firenze, and\\
Istituto Nazionale di Fisica Nucleare, Sezione di Firenze\\
Via G. Sansone 1,~I-50019~Sesto Fiorentino,~Firenze,~Italy}\\
$^2${\small\it  Frankfurt University of Applied Sciences\\
Nibelungenplatz 1,~D--60318~Frankfurt am Main,~Germany}}
%
\begin{abstract}The low--energy modes of a spin--imbalanced superfluid Fermi gas in
  the Bardeen-Cooper--Schrieffer (BCS) side are studied. The gas is assumed to
  be sufficiently dilute so that the pairing of atoms can be considered
  effective only in $s$--wave 	between fermions of different internal state. The order
  parameter at equilibrium is determined by the mean--field approximation,
  while the properties of the collective modes are calculated within a
  Gaussian approximation for the fluctuations of the order parameter. In
  particular we investigate the effects of asymmetry between the populations
  of the two different components and of temperature on the frequency and
  damping of collective modes. It is found that the temperature does not much
  affect the frequency and the damping of the modes, whereas an increase of
  the imbalance shifts the frequency toward lower values and enhances the
  damping sensitively. Besides the Bogoliubov--Anderson phonons, we observe
  modes at zero frequency for finite values of the wave--number. These modes
  indicate that an instability develops driving the system toward two separate
  phases, normal and superfluid.
  \end{abstract} 
\maketitle
\section{Introduction}
\label{intro}
Due to the addition of a further state parameter, the ratio between the
populations of the two different internal states, imbalanced two--component 
Fermi gases can display a very rich phase diagram with respect to the symmetric case. 
For instance, besides the continuous phase--transition, we can observe at
sufficiently low temperature a first--order transition between superfluid and
normal phases, driven by the imbalance parameter.  
Moreover, the possibility of tuning the strength of interatomic interactions
via Feshbach resonances allows to explore very different physical
regimes. Therefore, the studies of such systems involve multifaceted aspects
of many--body physics. These peculiar features have stimulated  several theoretical
works. By referring only to three--dimensional systems,  see
Refs. \cite{Ra10,Gu13}  for a general and extended review on the subject and
Refs. \cite{Bi14,Pa14,Bo15,Di15,Du16}  for more recent studies of the
phase structure of spin--imbalanced Fermi gases. In the meantime experimental
investigations have been performed in order to assess the occurrence of
different phases in trapped spin--imbalanced Fermi gases. More specifically,
phase separations have been indicated by analyzing the  density distributions
of two--component Fermi gases \cite{Zw06,Pa06,Pa06_1,Sh06,Sh08,Na10,Ol15}.   
\par
As the imbalance parameter increases, the Fermi gas experiences different physical 
situations, which can give rise to qualitatively new effects with respect to the 
balanced case. In particular the system can exhibit \cite{Gu13} a first order phase transition
and the onset of the gapless Sarma phase \cite{Sar63}. We should expect sizable effects 
of the degree of polarization on the dynamic behavior of the gas. 
In this paper we present a study on collective modes, having energy
below the pair--breaking threshold, of a polarized Fermi gas in the
Bardeen-Cooper-Schrieffer (BCS) regime. In particular, we are concerned with
the combined effects of the asymmetry of populations and temperature. The
parameters of the equilibrium state are determined within the saddle--point
approximation to the functional integral representing the grand partition
function of the system. For the collective modes we adopt a Gaussian
approximation for the time--dependent fluctuations of the order parameter
about its equilibrium value.  
\par
It is known that the saddle--point approximation or, equivalently, the
mean--field approach to the description of the properties of a superfluid
Fermi gas has only a qualitative validity when the unitary regime,
corresponding to resonant two--body interactions, is approached. This
limitation should be ascribed to the fact that the mean--field approximation
does not include thermodynamic and quantum fluctuations \cite{Gi08}. The
latter may have sizable effects in the region of the BCS--BEC
(Bose--Einstein--Condensate) crossover or when the values of temperature are
quite close to that of the normal--superfluid phase transition. There have
been calculations \cite{Di08,Kl11} taking into account the contribution of
the fluctuations of the order parameter within a framework similar to that
used here. However, the approach of Ref. \cite{Di08} suffers from a severe
drawback: quite generally, one can notice that the Goldstone theorem is
violated when the saddle--point equation is just modified away from its
mean--field expression \cite{Di08}. In view of that, we prefer to follow the
mean--field approximation to calculate the equilibrium order
parameter. Moreover, our calculations are limited to physical situations not
very close to the unitary regime or to the critical temperature of the
superfluid phase transition. Then, we may believe that our results are
reliable at least at a qualitative level.   
\par
The mismatch of the Fermi surfaces of the different species might give rise to
the inhomogeneous LOFF phase \cite{La64,Fu64}. However, experiments have found no
sign of such a phase in three--dimensional Fermi gases. So, its occurrence still
remains an open problem \cite{Zw06,Pa06,Pa06_1,Sh06,Sh08}. Here, we consider
only homogeneous phases.  

\section{Formalism}
We consider a two--component Fermi gas, whose components are only
distinguished by the internal quantum state, "spin up" ($+$) and "spin down"
($-$).  We consider  a  sufficiently dilute  gas so that we can assume that
the interatomic interaction has a short range and mainly occurs in the
$s$--wave channel.  Because of the Pauli principle the interaction can be
effective only between atoms of different species. 
\par
In the imaginary--time functional formalism an effective action for the
pairing field can be obtained by using a Stratonovich--Hubbard transformation
for the couples of fermion fields (see, e.g., Ref. \cite {Altland}). Within
this framework we derive the properties of the low--energy collective modes
taking into account Gaussian fluctuations of the pairing field about its
saddle point value. Here, we extend to the imbalanced case and to finite 
temperatures the formalism developed in Ref. \cite{Di08} for the Gaussian fluctuations. 
Further formal details can also be found in 
Ref. \cite{Mat15}. 
\par 
For the sake of completeness and in order to introduce notations, we report
the equation for the equilibrium pairing field $\Delta_0$  evaluated within
the saddle--point approximation to the effective action. A regularized
expression in terms of the scattering amplitude $a_S$ can be obtained with the 
substitution \cite{Eng97,Pet}  
\[\frac{1}{|\lambda|}\rightarrow -\frac{m}{4\pi a_S}+\int \frac{d{\bf k}}{(2\pi)^3}\frac{1}{2\epsilon(k)},\]
where $\lambda<0$ is the strength of a contact interaction between fermions and 
$\epsilon(k)=k^2/(2m)$ is the fermion kinetic energy. It reads 
\begin{eqnarray}
-\frac{m}{4\pi a_S}&&=
\int \frac{d{\bf k}}{(2\pi)^3}\Big[\frac{1}{4E^{(0)}(k)}
\label{gap}
\\
&&\times \bigg(\tanh(\frac{\beta E_+^{(0)}(k)}{2})+
\tanh(\frac{\beta E_-^{(0)}(k)}{2})\bigg)-\frac{1}{2\epsilon(k)}\Big],
\nonumber
\end{eqnarray}
where $\beta=1/T$, $E^{(0)}(k)=\sqrt{(\epsilon(k)-\mu)^2+\Delta_0^2}$  with $\mu=(\mu_++\mu_-)/2$
the average chemical potential,  and $E^{(0)}_{\pm}(k)=E^{(0)}(k)\mp\eta$ are the quasiparticle 
energies with $\eta=(\mu_+-\mu_-)/2$  half the difference in chemical potentials (units
such that $\hbar=k_B=1$ are used). 
\par    
Adapting the formal calculations of Refs. \cite{Di08,Mat15}  to the present case one can
derive, for the propagator of the pairing--field fluctuations \[\sigma({\bf
  q},\tau)=\Delta({\bf q},\tau)-\Delta_0,\]  
the $2\times 2$ set of algebraic  equations 
\begin{equation} 
{\widehat D}(\omega_n,{\bf q})=\frac{\lambda}{\cal V}{\widehat 1}
-\lambda\int \frac{d{\bf k}}{(2\pi)^3}
{\widehat A}(\omega_n,{\bf q},{\bf k})
{\widehat D}(\omega_n,{\bf q})\,,
\label{propg1}
\end{equation}
in the frequency representation, where $\omega_n$ are the Matsubara
frequencies and $\cal V$ is the normalization volume.  
The elements of the kernel ${\widehat A}(\omega_n,{\bf q},{\bf k})$ are
explicitly given by  
\begin{eqnarray}
&&A_{1,1}(\omega_n,{\bf q},{\bf k})=
\nonumber
\\
&&\big[1-f(E_-({\bf k}))-f(E_+^{\prime}({\bf k}))\big]
\frac{u^2({\bf k})u^{\prime 2}({\bf k})}{i\omega_n+E_-({\bf k})+E_+^{\prime}({\bf k})}
\nonumber
\\
&&-\big[1-f(E_-^{\prime}({\bf k}))-f(E_+({\bf k}))\big]
\frac{v^2({\bf k})v^{\prime 2}({\bf k})}{i\omega_n-E_-^{\prime}({\bf k})-E_+({\bf k})}
\nonumber 
\\
&&-\big[f(E_+^{\prime}({\bf k}))-f(E_+({\bf k}))\big]\frac{v^2({\bf k})u^{\prime 2}({\bf k})}
{i\omega_n +E_+^{\prime}({\bf k})-E_+({\bf k})}
\nonumber
\\&&-\big[f(E_-({\bf k}))-f(E_-^{\prime}({\bf k}))\big]\frac{u^2({\bf k})v^{\prime 2}({\bf k})}
{i\omega_n +E_-({\bf k})-E_-^{\prime}({\bf k})}
\label{a11}
\end{eqnarray}
and 
\begin{eqnarray}
&&A_{1,2}(\omega_n,{\bf q},{\bf k})=-\frac{\Delta_0^2}{4E({\bf k})E^{\prime}({\bf k})}\times\Big[
\nonumber
\\
&&\big[1-f(E_-({\bf k}))-f(E_+^{\prime}({\bf k}))\big]
\frac{1}{i\omega_n+E_-({\bf k})+E_+^{\prime}({\bf k})}
\nonumber
\\
&&-\big[1-f(E_-^{\prime}({\bf k}))-f(E_+({\bf k}))\big]
\frac{1}{i\omega_n-E_-^{\prime}({\bf k})-E_+({\bf k})}
\nonumber
\\
&&+\big[f(E_+^{\prime}({\bf k}))-f(E_+({\bf k}))\big]\frac{1}
{i\omega_n +E_+^{\prime}({\bf k})-E_+({\bf k})}
\nonumber
\\&&+\big[f(E_-({\bf k}))-f(E_-^{\prime}({\bf k}))\big]\frac{1}
{i\omega_n +E_-({\bf k})-E_-^{\prime}({\bf k})}\Big],
\label{a12}
\end{eqnarray}
while the remaining matrix elements are determined by the symmetry relations 
\[A_{2,2}(\omega_n,{\bf q},{\bf k})=A_{1,1}(-\omega_n,-{\bf q},{\bf k}),\]
\[A_{2,1}(\omega_n,{\bf q},{\bf k})=A_{1,2}(\omega_n,{\bf q},{\bf k}).\]
In the above equations $f(E_{\pm}({\bf k})$ and $f(E_{\pm}^{\prime}({\bf k})$
are the quasiparticle Fermi distributions and the following shorthand
notations are used: 
\[E_{\pm}({\bf k}),E_{\pm}^{\prime}({\bf k})=E({\bf k})\mp\eta,E^{\prime}({\bf k})\mp\eta,\]
\[u({\bf k}),v({\bf k}),E({\bf k})=u({\bf q}/2+{\bf k}), v({\bf q}/2+{\bf k}),E({\bf q}/2+{\bf k})\] 
and
\[u^{\prime}({\bf k}),v^{\prime}({\bf k}),E^{\prime}({\bf k})=u({\bf q}/2-{\bf k}), 
v({\bf q}/2-{\bf k}),E({\bf q}/2-{\bf k}),\]
where  $E({\bf q}/2\pm{\bf k})=\sqrt{\xi^2({\bf q}/2\pm{\bf k})+\Delta_0^2}$, 
\[u^2({\bf q}/2\pm{\bf k})=\frac{1}{2}\Big(1+\frac{\xi({\bf q}/2\pm{\bf k})}{E({\bf q}/2\pm{\bf k})}\Big),\]
and
\[v^2({\bf q}/2\pm{\bf k})=\frac{1}{2}\Big(1-\frac{\xi({\bf q}/2\pm{\bf k})}{E({\bf q}/2\pm{\bf k})}\Big),\] 
with $\xi({\bf q}/2\pm{\bf k})=({\bf q}/2\pm{\bf k})^2/2m-\mu$. 
\par
The propagator ${\widehat D}(\omega_n,{\bf q})$ represents the effective
interaction between quasi--particles when the exchange of phonons is
included. It is formally similar to the random phase approximation (RPA) to
the in medium effective interaction for normal Fermi systems \cite{Negele}.   
The Green`s functions of the pairing field  
${\widehat\sigma} ({\bf q},\tau)=[\sigma^* ({\bf q},\tau),\sigma({\bf
  q},\tau)]^\dag$ are obtained by subtracting from ${\widehat D}(\omega_n,{\bf
  q})$ the matrix element, in the momentum representation,  
$\lambda/{\cal V}$ of the bare interaction,  then by amputating the two
external interaction vertices:  
\[
\frac{{\cal V}^2}{\lambda^2} \bigg({\widehat D}(\omega_n,{\bf q}) -\frac{\lambda}{\cal V}{\widehat 1}\bigg) .\]
From Eqs.  (\ref{propg1}) we get an equivalent set of equations for the
propagator of the pairing fluctuations, ${\widehat\sigma} ({\bf q},\tau)$, per
unit of volume,  
\begin{equation}
{\widehat {\cal D}}(\omega_n,q)=-\frac{1}{|\lambda|}\big[\frac{\widehat 1}{|\lambda|}-
{\widehat A}^T(\omega_n,q)\big]
^{- 1}{\widehat A}^T(\omega_n,q)
\label{prop}
\end{equation}
with
 \[{\widehat A}^T(\omega_n, q)=\int \frac{d{\bf k}}{(2\pi)^3}{\widehat A}(\omega_n,{\bf q},{\bf k})\]
 depending only on the magnitude of the wave--vector ${\bf q}$. 
\par
The poles of the Green`s functions are given by the complex frequencies $z=\omega(q)+i\Gamma(q)$
 for which the determinant 
 \begin{equation}
\det\big[\frac{{\widehat 1}}{|\lambda|}- {\widehat A}^T(z,q)\big]
\label{det}
\end{equation} 
vanishes. In the integrals for the diagonal elements $A_{i,i}^T$,  the ultraviolet
divergencies are removed by substituting the coupling constant $\lambda$  with
the scattering length as in the case of the gap equation (\ref{gap}).  
\par
For given values of the wave--number $q$ the real part of the frequency
represents the excitation energy of collective modes while the imaginary part
gives their life--time. Then, we expect complex poles in the lower half--plane of $z$, $\Gamma(q)<0$, 
corresponding to damped oscillations. Since the Green`s functions have a branch cut along the real 
axis $\omega$, they should be analytically continued from the upper to the lower half--plane of 
$z$ on another Riemann sheet \cite {Noz,Pin}. In our case this can be obtained by adding 
$2ImA^T_{i,j}(\omega+i\epsilon,q)$ to the advanced counterpart $A^T_{i,j}(\omega-i\epsilon,q)$, which is in turn analytic in the lower half--plane. In the Appendix the procedure is exemplified by explicit  calculations in the limit of small wave--number $q$. 
Complex frequencies, occurring even for low values of
$\omega$ and $q$, emerge from the quasiparticle--quasihole singularities of
the last two terms of Eqs. (\ref{a11})and (\ref{a12}).  We stress in passing that
the factors $\big[f(E_{\pm}({\bf k})-f(E_{\pm}^{\prime}({\bf k})\big]$ vanish
only for $T=0$. Then, an expansion at small values of $q$ and $\omega$ is not
allowed in general \cite{Abr66}. Moreover, we observe that  
the onset, for sufficiently high imbalance $|\eta|\geq |\Delta_0|$ , of the gapless Sarma phase
does not give rise to any critical behavior of the properties of collective modes, 
since in the denominators of Eqs. (\ref{a11}) and (\ref{a12}) the energies of the intermediate states do not depend on $\eta$: $E_{\pm}^{\prime}({\bf k})+E_{\mp}({\bf k})=E^{\prime}({\bf k})+E({\bf k})$ and 
$E_{\pm}^{\prime}({\bf k})-E_{\pm}({\bf k})=E^{\prime}({\bf k})-E({\bf k})$.   
\section{Results}
\label{sec:3}
In this paper we are concerned with the properties of low--energy collective
modes, therefore we choose sufficiently low values of the wave--vector $q/k_{F_+}\leq 0.1$, 
where $k_{F_+}$ is the Fermi momentum of the spin--up fermions,
so that the corresponding energies are below the threshold for pair breaking. 
\par
To be specific, from now on we assume that the majority particles are in the
intrinsic state "spin--up", $n_+\geq n_-$, where $n_{\pm}$ are the densities
of the two species of fermions. We note that for $T\rightarrow 0$ the 
polarization $(n_+-n_-)$ is vanishing necessarily \cite{Gu13,Bar10}.  
\par
\begin{figure}
\includegraphics{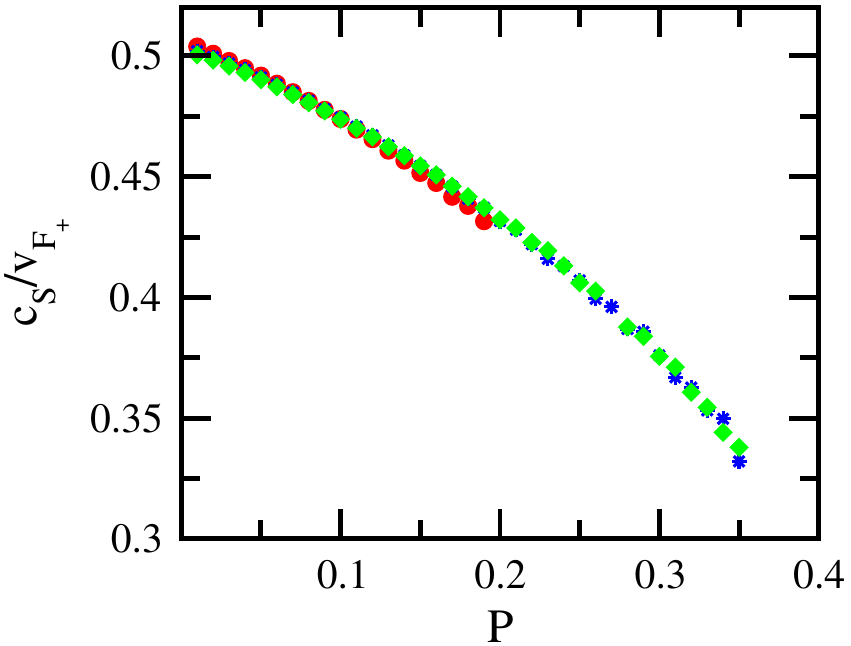}
\caption{ The scaled phase velocity of phonon--like modes as a function of the
  relative polarization for three values of temperature:
  $T=0.08\epsilon_{F_+}$ (red circles). $T=0.109\epsilon_{F_+} $(blue stars) and
  $T=0.12\epsilon_{F_+}$ (green diamonds). Calculations are done with $
  a_Sk_{F_+}=-2$.} 
\label{fig:1}
\end{figure}
\begin{figure}
\includegraphics{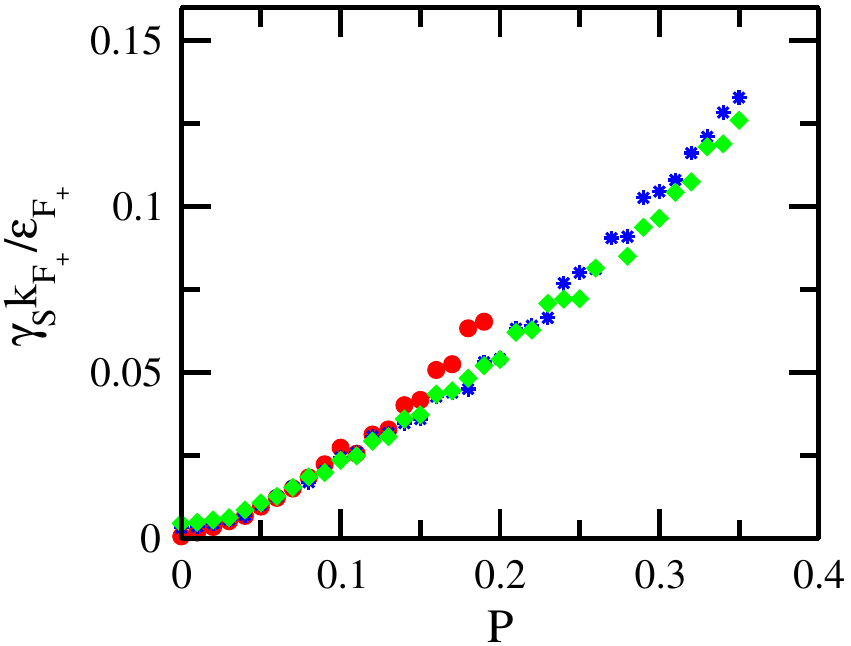}
\caption{ Same as Fig.1,  but for the scaled damping rate (see text) of phonon--like modes.}
\label{fig:2}
\end{figure}
In order to explore a fairly large region of the phase diagram $(P,T)$ accessible by the system, 
with $P=(n_+-n_-)/(n_++n_-)$ the relative polarization, we use the value
$a_Sk_{F_+}=-2$ for the interaction strength in actual calculations
\cite{Gu13,Bar10}. Furthermore, with this 
value of $a_S$ the physical parameters of the system are not close to the area
of the BCS--BEC cross--over, where fluctuations of the pairing field could be
important and our approach can be rather questionable.   
\par
For given values of the temperature we have calculated the frequency and the
life--time of collective modes as functions of the polarization, up to
values of $P$ near the line of the first--order  phase transition, for
$T<T_{CP}$, or the second--order  phase transition, for $T<T_{CP}$. Here
$T_{CP}$ is the temperature of the tricritical point, where the normal state,
the superfluid state and the instability region meet \cite{Gu13,Bar10}. We
have found that, besides a phonon--like pole, the propagator of the pairing
field, $ {\widehat {\cal D}}(z,{\bf q})$, exhibits a pole at $z\rightarrow 0$
for finite values of $q$, when $P$ approaches the border of the instability
region ($T<T_{CP}$). In particular, the pole moves along the imaginary axis
from the lower part to the upper part of the $z$--plane. The values of the
couple ($T,P$), for which the pole emerges, are in agreement with the
instability curve obtained in Refs. \cite{Gu13,Bar10} within a
thermodynamical approach. Finally, we note that for a given value of $T$ the
position of the pole in the $(P,T)$ plane moves to the right increasing $q$,
albeit slightly for the considered values of $q$. Similar features have been
found for the relevant response functions in investigations of first--order
phase transitions within different contexts \cite{Abr79,Mat00,Cho04}.  
\par 
The phase velocity $c_S=\omega(q)/q$ and the damping rate
$\gamma_S=|\Gamma(q)|/q$ of the phonon--like mode are displayed in Figs. (1)
and (2) respectively, as functions of the relative polarization for three
values of the temperature: below, at and above the tricritical temperature
($T_{CP}=0.109T_{F_+}$), with $T_{F_+}$ the Fermi temperature of the majority
particles. The curves end for values of $P$ at the border of the instability
region ($T=0.08T_{F_+}$), at the tricritical point ($T=0.109$) and at the
border of the second-order phase transition ($T=0.12T_{F_+}$).  
Here we have used the value $q=0.01k_{F+}$ for the wave--vector. However,
explicit calculations show that for the considered values of $q$ 
both the frequency and the life--time of the phonon--like excitations are
linear functions of $q$, with a good approximation. 
\par 
One can see from Figs. (1) and (2) that the phase velocity as well as the
damping rate show a sizable dependence on the polarization, whereas, in the
interval of $P$ where the collective modes subsist, they are slightly
affected by the temperature. In particular, the growing with $P$ of the
damping rate $\Gamma(q)/q$ can essentially be ascribed to the increase of the
density of quasi--particle states  
\[4\pi k^2\frac{\partial}{\partial k}\big[f(E_+(k))+f(E_-(k))\big] \]
with increasing the difference in chemical potentials, $\mu_+-\mu_-$. 
We note that the damping assumes appreciable values at relatively low temperature for 
sufficiently high values of $P$. Whereas for a balanced gas it 
becomes significant only at temperatures close to 
that of the normal--superfluid transition ( $T\sim 0.9T_C$ ) \cite{Oh03,Ur06}. 
\par
\begin{figure}
\includegraphics{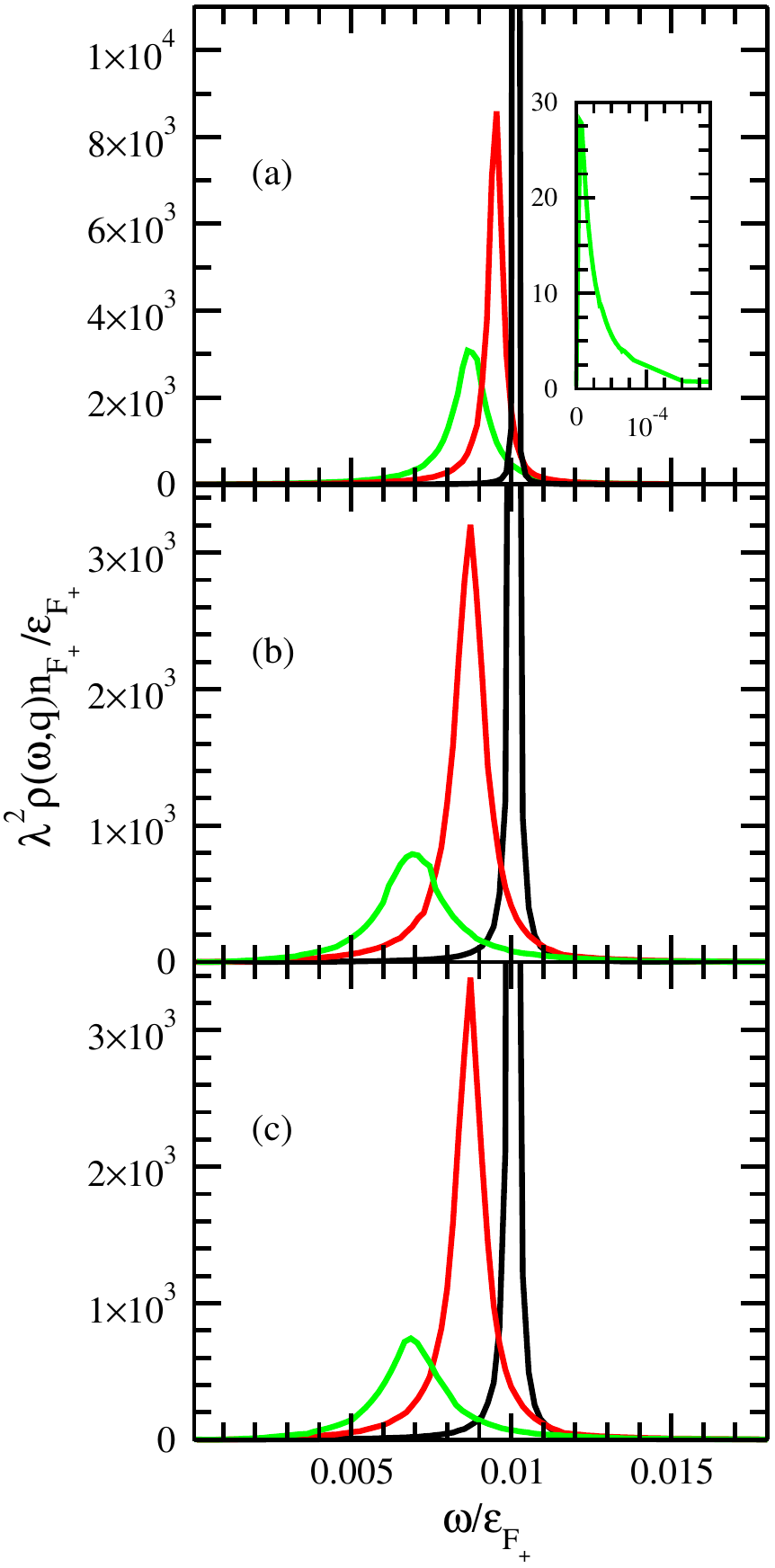}
\caption{ The modified spectral density of collective modes (see text) as
  function of the scaled frequency $\omega/\epsilon_{F_+}$ for different
  values of temperature and relative polarization. Panel(a):
  $T=0.08\epsilon_{F_+}$, $P=0$ (black line), $P=0.1$ (red line) and $P=0.19$
  (green line). The inset in panel (a) displays a zoom on low--frequency
  region for $P=0.19$. Panels (b) and (c) show  the results for
  $T=0.109\epsilon_{F_+}$ and $T=0.12\epsilon_{F_+}$ respectively, with three
  values of polarization, $P=0.0$ (black line), $P=0.2$ (red line) and
  $P=0.33$ (green line). The scaled wave--number is $q/k_{F+}=0.01$ and the
  used value for $ a_Sk_{F_+}$ is the same as in Fig.1.  
}
\label{fig:3}
\end{figure} 
\begin{figure}
\includegraphics{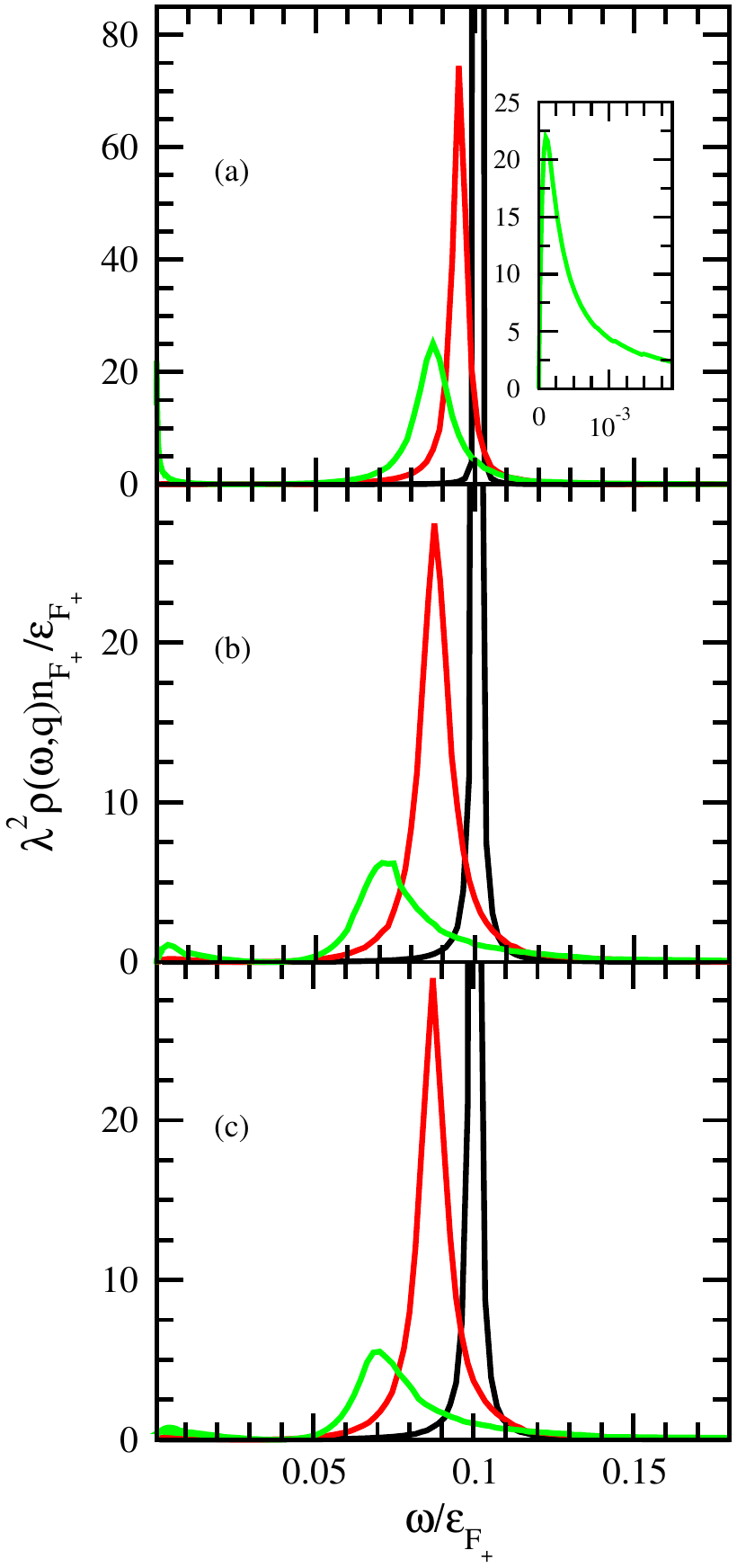}
\caption{ Same as Fig.3,  but for $q/k_{F_+}=0.1$.}
\label{fig:4}
\end{figure}
The spectral density of the pairing--field fluctuations can supply a more
detailed insight of the low--energy excitation spectrum of the superfluid
phase. This quantity corresponds to the  imaginary part of the retarded
propagator of the pairing--field fluctuations ${\widehat\sigma} ({\bf
  q},\tau)$, and can be obtained by analytical continuation of the matrix
element ${\cal D}_{1,1}(\omega_n,q)$ to real frequencies (
$i\omega_n\rightarrow \omega+i\epsilon$ ) 
\[
\rho(\omega,q)=-\frac{1}{\pi}Im{\cal D}_{1,1}(\omega,q) .\]
It is convenient to rewrite Eqs. (\ref{prop}) as 
\begin{equation}
{\widehat {\cal D}}(\omega,q)=-\frac{1}{\lambda^2}\Big(\big[\frac{\widehat 1}{|\lambda|}-
{\widehat A}^T(\omega,q)\big]
^{- 1}-|\lambda|{\widehat 1}\Big). 
\label{prop1}
\end{equation}
Thus, the spectral density is given by 
\begin{equation}
\rho(\omega,q)=\frac{1}{\lambda^2}\frac{1}{\pi}Im\big[\frac{\widehat 1}{|\lambda|}-
{\widehat A}^T(\omega,q)\big]_{1,1}^{- 1},
\label{spect}
\end{equation} 
and for the integrals we can exploit the regularization procedure used before,
Eq. (\ref{gap}). The pre--factor $1/\lambda^2$ in the above equation is a
simple scale factor non affecting the relevant features of
$\rho(\omega,q)$. We can safely omit it and limit ourselves to examine the
quantity $\lambda^2\rho(\omega,q)$. In any case the coupling constant
$\lambda$ could be determined consistently with the value of the scattering
length $a_S$. We note, however, that the relationship between $\lambda$ and
$a_S$, used in the present work, is not suitable for this purpose, we should
introduce a further parameter at least.  
\par
In Figs. (3) and (4) we show the calculated spectral density as a function of
$\omega$ for two values of the wave--vector, $q=0.01k_{F+}$ and $q=0.1k_{F+}$
respectively. The values of temperature are the same as in Figs. (1) and (2):
below, at, and above the tricritical temperature. For each temperature three
values of the relative polarization $P$ are considered, the last of which
being in proximity of the curve of the first or second order
phase--transition. The inset in the figures is a zoom on the region about
$\omega =0$, showing the occurrence of the instability pole. Furthermore, in
the graphs with $T=0.08T_{F_+}$, we have chosen the highest value of $P$ taking
into account that the instability border is shifted toward larger  
$P$ increasing the value of $q$.  
\par
We observe that in the two figures abscissae and widths of the peaks only differ
by a scale factor, given by the ratio between the values of  wave vector. This
denotes a linear dependence on the wave vector of the complex frequency, as
mentioned before. In addition, Figs. (3) and (4) show that the global
properties of the spectral density depend mostly on polarization, whereas the 
temperature, in the considered range, has a weak impact. 
A remark is in order. For low values of the polarization an undamped hydrodynamic sound mode 
can survive until temperatures comparable to the critical one \cite{Ur06}. Increasing the polarization 
the microscopic mechanism underlying the collective mode becomes different. 
The quasiparticle--quasihole  states acquire a more and more important weight, yielding an 
anisotropic distortion of the equilibrium quasiparticle distribution, so that the requirement of  
local equilibrium for a hydrodynamic sound mode could no longer be   satisfied.  
\par
We finish the survey on the properties of the spectral density giving a look
at its composition in terms of amplitude and phase modes. It is convenient to
express the pairing--field fluctuations as  
$\sigma({\bf q},\tau)=\chi({\bf q},\tau)+i\phi({\bf q},\tau)$, where
$\chi({\bf q},\tau)$ and $\phi({\bf q},\tau)$ are real and may be identified
with amplitude and phase fluctuations, respectively. Their propagators are
given by simple combinations of the matrix elements of ${\widehat {\cal
    D}}(\omega,{\bf q})$:  
\begin{eqnarray}
{\cal D}_{\chi}(\omega,{\bf q})=&&\frac{1}{4}\big[ {\cal D}_{1,1}(\omega,{\bf
  q})+{\cal D}_{2,2}(\omega,{\bf q})
  \nonumber
  \\
&&+{\cal D}_{1,2}(\omega,{\bf q}) +{\cal
  D}_{2,1}(\omega,{\bf q}) \big] 
  \nonumber
 \end{eqnarray} 
and
\begin{eqnarray}
{\cal D}_{\phi}(\omega,{\bf q})=&&\frac{1}{4}\big[ {\cal D}_{1,1}(\omega,{\bf
  q})+{\cal D}_{2,2}(\omega,{\bf q})
  \nonumber
  \\
  &&-{\cal D}_{1,2}(\omega,{\bf q})-{\cal
  D}_{2,1}(\omega,{\bf q}) \big] . 
\nonumber
\end{eqnarray}
Explicit calculations of the imaginary parts of these quantities show that,
for both the considered values of $q$, the phase and amplitude modes are
almost decoupled: the phonon peak corresponds to a damped phase mode ( $>90\%$
) in all the cases, whereas the instability pole ( $T<T_{CP}$ ) corresponds to
an unstable amplitude mode ( $>90\%$ ).

\section{Summary and conclusions}
\label{summary}
In our approach to assess the occurrence of a superfluid phase and to
determine the properties of low--energy excitations of a imbalanced
two--component Fermi gas, only fermionic degrees of freedom come into
play. For a sufficiently diluted gas only atoms, in different "spin" states
and in the $s$--channel, can be considered paired.  
\par
So far, imbalanced Fermi gases have been studied at equilibrium mostly.  A
particular attention has been devoted to their properties in the unitarity
regime. Here, we have looked into some aspects of the dynamics, which may
play a role in the response of the gas to an external probe. We have focused
our attention into the effects of the polarization of the gas on the
low--energy excitation spectrum. The effects of the temperature have been
taken into account as well. At finite temperatures, singularities, due
quasiparticle excitations, emerge in the integrals. They give rise to damped
collective modes, like the well--known Landau damping. In the case of a
balanced Fermi gas, a strong increase of the width of the peak, corresponding
to the Bogoliubov--Anderson mode, was observed  for values of temperature
close to the critical one \cite{Oh03,Ur06}. In our study
we have found a similar behavior but with increasing the polarization of the
gas, even at temperatures far from the critical one. Moreover, we have not
observed any critical effect on the low--lying collective modes from the
occurrence of the Sarma phase \cite{Sar63}.  Actually, approaching the Sarma
phase the density of quasiparticle states increases but continuously.   
\par
Our calculations show that, in addition to the pole related to the
Bogoliubov--Anderson mode, the propagator of the pairing field exhibits a
purely imaginary pole. The magnitude of the imaginary frequency is vanishing
when the value of the polarization reaches the borders of the coexistence
region of the superfluid and normal phases.  This is connected to an
instability situation, which develops toward the separation of the two
phases. However we remark that, in the formalism used in the present paper,  
the normal phase, when the order parameter vanishes, reduces to an ideal
gas. In order to attain a more realistic description of the phase separation,
the formalism should be extended to include an interaction between
quasiparticles.     
\appendix
\section*{Appendix}
\setcounter{equation}{0}
\section{Analytic continuation}
\numberwithin{equation}{section}
For excitation energies below the pair--breaking threshold the terms in Eqs. (\ref{a11}) and (\ref{a12}) 
giving rise to complex poles of the Green`s function of Eq. (\ref{prop}) are those which contain the 
factors  $\big[f(E_{\pm}({\bf k})-f(E_{\pm}^{\prime}({\bf k})\big]$. For small wave numbers $q$ 
( $qv_{F_+}<<\Delta_0$ ), their contribution to $A^T_{i,j}(\omega_n,q)$ is the same for all 
the matrix elements. It is given by 
\begin{eqnarray}
\delta A^T(\omega+i\epsilon,&&q)=-\int \frac{d{\bf k}}{(2\pi)^3}u_0^2(k)v_0^2(k)
\label{a1}
\\
\nonumber
&&\times 2\bigg(\frac{\nabla_{\bf k}f(E^{(0)}_+(k))\cdot{\bf q}}{\omega+i\epsilon+2{\bf q}\cdot{\bf w}}
-\frac{\nabla_{\bf k}f(E^{(0)}_-(k))\cdot{\bf q}}{\omega+i\epsilon-2{\bf q}\cdot{\bf w}}\bigg),
\end{eqnarray}
where the retarded case ( $i\omega_n\rightarrow \omega+i\epsilon$ ) is considered. In the above equation the notations ${\bf w}=\nabla_{\bf k}E^{(0)}(k)$ and $u_0(k),v_0(k)=u({\bf k}),v({\bf k})$, 
with $q=0$, are used ( see Eqs. (\ref{a11}) and (\ref{a12}) ). \par
The angular integration in Eq. (\ref{a1}) can be performed analytically and gives 
\begin{eqnarray}
\delta A^T(\omega+i\epsilon,q)&&=\int \frac{dk}{(2\pi)^2}k^2\frac{u_0^2(k)v_0^2(k)}{T}
\label{a2}
\\
\nonumber
&&\times\bigg[\frac{e^{\beta E^{(0)}_+(k)}}{\big(1+e^{\beta E^{(0)}_+(k)}\big)^2}+
\frac{e^{\beta E^{(0)}_-(k)}}{\big(1+e^{\beta E^{(0)}_-(k)}\big)^2}\bigg]
\\
\nonumber
&&\times\bigg[2-\frac{\omega}{2qw}\ln\big(\frac{\omega+2qw+i\epsilon}{\omega-2qw+i\epsilon}\big)
\bigg].
\end{eqnarray}
The analytic continuation to the lower half--plane of the complex frequency can be made by exploiting the property that in the complex plane the logarithm is a multi--valued function. The logarithms of the advanced counter--part, $\delta A^T(\omega-i\epsilon,q)$, are evaluated on the next Riemann sheet: 
\begin{eqnarray}
ln(\omega-i\epsilon-2qw)&&=ln(|\omega-i\epsilon-2qw|)
\nonumber
\\
&&+i\big(\arg(\omega-i\epsilon-2qw)+2\pi\theta(1-s)\big),
\nonumber
\end{eqnarray}
for $s=\omega/2qw>0$, or
\begin{eqnarray}
ln(\omega-i\epsilon-2qw)&&=ln(|\omega-i\epsilon-2qw|)
\nonumber
\\
&&+i\big(\arg(\omega-i\epsilon-2qw)+2\pi\theta(1+s)\big),
\nonumber
\end{eqnarray}
for $s=\omega/2qw<0$. This is equivalent to defining the analytic continuation of the retarded function 
as \[\delta A^{T\prime}(\omega-i\epsilon,q)=\delta A^T(\omega-i\epsilon,q)+2iIm\delta A^T(\omega+i\epsilon,q).\]

\end{document}